\documentclass[submission,copyright,creativecommons]{eptcs}
\providecommand{\event}{FMAS 2023} 
\usepackage{iftex}
\usepackage{underscore}               
\usepackage[utf8]{inputenc}
\usepackage[T1]{fontenc}
\usepackage{comment}
\usepackage{bookmark}          
\usepackage{url}               
\usepackage[all]{foreign}      
\usepackage{booktabs}          
\usepackage{cite}
\usepackage{multicol}
\usepackage{marginnote}        
\usepackage[only,llbracket,rrbracket]{stmaryrd} 
\usepackage[inline]{enumitem}
\usepackage{amsmath,amssymb, amsthm}   
\usepackage{todonotes}
\usepackage{thmtools, thm-restate}
\usepackage{graphicx}          
\usepackage{listings}
\usepackage{multirow}

\lstdefinelanguage{smv}{
  basicstyle=\scriptsize\ttfamily,
  keywords = {MODULE,INIT,TRANS,INVAR,VAR,IVAR,next,integer,INVARSPEC,LTLSPEC,F,G,boolean,DEFINE},
  aboveskip={0\baselineskip},
  frame=single,
  columns=fixed,
  keywordstyle=\bfseries\ttfamily\color{blue},
  commentstyle=\it\ttfamily\color{red},
  prebreak = \raisebox{0ex}[0ex][0ex]{\ensuremath{\hookleftarrow}},
  showstringspaces=false,
  showtabs=false,
  showspaces=false,
  breaklines=false,
  extendedchars=true,
  numbers=none,
  numberstyle=\tiny,
  numbersep=5pt,
  stepnumber=1,
  classoffset=1, 
  otherkeywords={TODO},
  morekeywords={TODO},
  keywordstyle=\color{red},
  classoffset=0,
  morecomment=[l]{--}
}

\newcommand{\carla}{CARLA\xspace}
\newcommand{\scenariorunner}{ScenarioRunner\xspace}

\title{Automatic Generation of Scenarios for System-level Simulation-based Verification of Autonomous Driving Systems\thanks{This work has been supported by: the “AI@TN” project funded by the Autonomous Province of Trento; the PNRR project FAIR - Future AI Research (PE00000013), under the NRRP MUR program funded by the NextGenerationEU; and the PNRR MUR project VITALITY (ECS00000041), Spoke 2 ASTRA - Advanced Space Technologies and Research Alliance.}}

\author{Srajan Goyal
\institute{Fondazione Bruno Kessler\\ University of Trento\\ Trento, Italy}
\email{sgoyal@fbk.eu}
\and
Alberto Griggio
\institute{Fondazione Bruno Kessler\\ Trento, Italy}
\email{griggio@fbk.eu}
\and
Jacob Kimblad
\institute{Fondazione Bruno Kessler\\ Trento, Italy}
\email{jkimblad@fbk.eu}
\and
Stefano Tonetta
\institute{Fondazione Bruno Kessler\\ Trento, Italy}
\email{tonettas@fbk.eu}
}
\def\titlerunning{Automatic Generation of Scenarios for ADS}
\def\authorrunning{Goyal, Griggio, Kimblad, and Tonetta}

\begin{document}
\maketitle

\begin{abstract}
  With increasing complexity of Automated Driving Systems (ADS),
  ensuring their safety and reliability has become a critical
  challenge. The Verification and Validation (V\&V) of these systems
  are particularly demanding when AI components are employed to
  implement perception and/or control functions.
  In ESA-funded project VIVAS, we developed a generic framework
  for system-level simulation-based V\&V of autonomous systems. The
  approach is based on a simulation model of the system, an abstract
  model that describes symbolically the system behavior, and formal
  methods to generate scenarios and verify the simulation
  executions. Various coverage criteria can be defined to guide the
  automated generation of the scenarios.

  In this paper, we describe the instantiation of the VIVAS framework
  for an ADS case study. This is based on the integration of \carla, a
  widely-used driving simulator, and its \scenariorunner tool, which
  enables the creation of diverse and complex driving scenarios. This
  is also used in the \carla Autonomous Driving Challenge to validate
  different ADS agents for perception and control based on AI, shared
  by the \carla community. We describe the development of an abstract
  ADS model and the formulation of a coverage criterion that focuses
  on the behaviors of vehicles relative to the vehicle with ADS under
  verification. Leveraging the VIVAS framework, we generate and
  execute various driving scenarios, thus testing the capabilities of
  the AI components. The results show the effectiveness of VIVAS in
  automatically generating scenarios for system-level simulation-based
  V\&V of an automated driving system using \carla and
  \scenariorunner. Therefore, they highlight the potential of the
  approach as a powerful tool in the future of ADS V\&V methodologies.

\end{abstract}

\section{Introduction}

In the rapid evolution of Autonomous Driving Systems (ADS), the
problem of ensuring their safety and reliability has become a
paramount concern. The Verification and Validation (V\&V) of these
systems necessitate the assessment of their correctness in a multitude
of dynamic and complex real-world scenarios. To address this
challenge, the integration of powerful simulation tools with advanced
verification methodologies has gained considerable attention
\cite{verifai-cav19,paracosm}.

Under the support of ESA funding, the VIVAS project\cite{vivas} was dedicated to
developing a generic framework tailored for system-level simulation-based
V\&V of autonomous systems. The approach is based on a simulation
model of the system, an abstract model that describes symbolically the
system behavior, and formal methods to generate scenarios and
verify the simulation executions. It permits the specification of
diverse coverage criteria, thereby directing the automated creation of
scenarios, and formal properties to be verified on the
simulation runs. The framework has been created for space applications
and applied to two use cases employing AI for resource
prediction and opportunistic science. The system under test is based on the robotic digital
twin developed in ROBDT~\cite{robdt}. 

In the automotive context, the \carla simulator~\cite{carla} has
established itself as a widely used platform for simulating intricate
driving scenarios in a controlled virtual environment. Its
\scenariorunner tool~\cite{scenariorunner} further enhances its
capabilities by enabling the specification of diverse and complex
scenarios based on the reuse of various predefined car behaviors.
Various works proposed AI-based solutions (e.g.,
\cite{shao2022safetyenhanced,wu2022trajectoryguided,chen2022learning})
for the perception and control components of cars that are integrated
with the \carla simulator for their validation. The \carla community
also organized a competition to compare and rank such
solutions~\cite{carla-leaderboard}. These autonomous driving agents,
grounded in AI methodologies, serve as crucial components in achieving
the autonomy of cars. Their integration in \carla allows evaluating the
ADS behavior across different scenarios. However, such validation is
so far based on a few manually crafted scenarios.

This paper explores the application of the VIVAS framework to
automatically generate scenarios for a system-level
simulation-based verification of autonomous driving systems. This
integration facilitates a comprehensive assessment of ADS correctness
under various conditions, contributing to the enhancement of their
safety and reliability. The paper describes the abstract ADS model in
the extended SMV language handled by the nuXmv~\cite{nuxmv} symbolic model checker, capturing essential aspects of its functionality and behavior in
a simple highway traffic situation. It then details the formulation of
a coverage criterion based on such an abstract model, focusing on the
interactions between other vehicles and the vehicle of the ADS under verification (hereafter called \textit{ego}). The VIVAS integration finally
consists of a translation of the abstract traces generated from the
abstract model to the \scenariorunner specification and a mapping back
of the simulation runs to abstract traces for runtime verification of
formal properties. The experimental evaluation shows how VIVAS is able
to generate interesting scenarios effectively evaluating the behavior
of the AI-based agents.

The rest of the paper is organized as follows: in
Section~\ref{sec:rw}, we summarize the related works and compare them
with our approach; Section~\ref{sec:vivas} describes in more
detail the VIVAS framework and its components; in
Section~\ref{sec:ads}, we detail the instantiation to the ADS
application; Section~\ref{sec:results} shows the results while
Section~\ref{sec:conclusions} draws conclusions and some
directions for future work.

\section{Related Work}
\label{sec:rw}

Over the last decade, we have witnessed significant efforts in the
verification of AI-based autonomous systems using formal methods. Many
works focus on formal verification of neural networks, for example
encoding them into constraint solving (e.g.,
\cite{HuangKWW17,reluplex,marabou}) or using abstraction
(e.g., \cite{SinghGPV19,PasareanuMGYICY23}), just to name a few
approaches. Our approach instead is rooted in the line of research (e.g., \cite{pegasus,sim-atav,verifai-cav19}) that tackles the
verification at the system level using a simulator. This integrates the AI
components, potentially using machine learning (ML) models, for perception or control,
in the context of a closed-loop cyber-physical system. As in
VerifAI~\cite{verifai-cav19}, the simulation traces are then formally
analyzed with monitoring and runtime verification techniques.

Differently from the mentioned approaches, we exploit an abstract
symbolic model to generate automatically the scenarios and define a
coverage criterion for the generated test cases. While previous
approaches focus on the automated synthesis of the simulation
parameters for a specific scenario 
(e.g.,
different car movements to change lanes in front of the ego car), we
concentrate on the generation of different functional scenarios (e.g.,
sequences of scenes with different change lanes of non-ego
cars). Moreover, in this paper, we map such abstract symbolic scenarios
to the scenario specification language of \carla to verify ADS with
different available AI solutions. So, the case study is based on
available benchmarks for AI-based ADS taken as is.

There are in fact a variety of scenario specification languages that
can be used in this context.
VerifAI uses the Scenic language~\cite{Fremont2022,VinKRFKDGYSS23} to model the
abstract feature space defining the scenarios, which can be
instantiated to test cases. Scenic is a probabilistic programming
language for scenario generation specifically designed to test the
robustness of systems containing AI and ML components by allowing the
generation of rare events. It allows the specification of spatial
and temporal relationships between objects of a scenario as well as
composing several scenarios into more complex ones. By the
use of distributions for encoding interesting parameters, Scenic will
perform automatic test case generation through the use of sampling.
Similarly, the Paracosm\cite{paracosm} framework is a programmatic
interface that can be used to create various automotive driving
simulation scenarios through the design of parameterized environments
and test cases. The parameters control the environment in the scenario
including the behavior of the actors and can include things such as
pedestrians, lanes, and light conditions. Parameters are specified
using either discrete or continuous domains and test cases are
instantiated from the domain using random sampling and Halton sampling
respectively. A coverage criterion is then defined over the coverage of
the domains, where k-wise combinatorial coverage is used for the
discrete domains and dispersion is used for continuous
domains. Although Paracosm can provide output using the OpenDRIVE
format, it is primarily coupled to be used with the Unity game engine,
and as such scenarios are modeled using the C\# programming language.
The Measurable Scenario Description Language (M-SDL)\cite{m-sdl} is another scenario
description language similar to Scenic. In M-SDL, one captures the behavior of
identified actors in scenarios. M-SDL makes use of pre-defined basic building
blocks such as actors (including the AV) along with some pre-defined behaviors,
sets of possible routes, and environmental conditions. Libraries then use the
basic building blocks to implement more complex behaviors such as cars
overtaking, running red lights, driving on a highway, etc. Since M-SDL
scenarios are abstract and parameterized, a single scenario can map onto many
concrete ones through the use of sampling.
\scenariorunner~\cite{scenariorunner} is a module of \carla that allows
traffic scenario definition and execution for the \carla simulator. The
scenarios can be defined through a Python interface or using the
OpenSCENARIO standard~\cite{openscenario}. \scenariorunner is used to
validate AI solutions for ADS. These results can be validated and
shared in the \carla Autonomous Driving Leaderboard~\cite{carla-leaderboard}, an open
platform for the community to fairly compare their progress,
evaluating agents in realistic traffic situations.

For all the above languages, the scenario must be specified manually,
to then derive the test cases automatically. VIVAS instead provides
a model-based approach to generate the scenarios automatically based on
a coverage criterion that defines the interesting combinations of
situations. In this paper, we focus on the integration with \carla,
because it allows the verification of the solutions shared by the
\carla community. However, the approach can also work with different
specification languages, and we have, for example, a prototype integration
with Scenic interfaced with \carla. We have not presented the results of this integration in this paper since the ego model is based on Newtonian physics, with no AI models involved in the autonomous driving pipeline.

Although not specifically focused on AI-based systems, another very
relevant work is described in \cite{KlischatA20}, which proposes an
optimization-based approach to synthesize ADS scenarios from
formal specifications and a given map. Their formal specification of
scenarios corresponds to our abstract scenario and is also synthesized
from a symbolic model. However, test case generation does not follow any
coverage criteria but enumerates specifications starting from an
initial scene.
In principle, our coverage-driven generation of
scenarios can be combined with various techniques to concretize the
scenario with different trajectories and sampling of the different
environment parameters.

TAF \cite{taf} is another tool for automated test case generation of
autonomous systems. Their abstract model is defined in an XML-based
domain-specific language. It includes semantic constraints on the
initial conditions of the environment and its agents (unlike the
additional state transition systems in our work), which are solved
using SMT solvers to generate abstract test cases. Random sampling is
combined with these solvers to diversify the test cases, with an
expert given coverage of data values. Their coverage criteria is based
on covering parameter values to instantiate the scenarios. Although
constraints on time can be expressed, more generic temporal
specification on the sequences of actions and the related coverage
criteria are not supported as in our approach. On the other side, our
framework can be extended to constraints with quantifiers and complex
data structure as in~\cite{SartoriWG23}, which are currently not
supported in VIVAS.

\section{The VIVAS Framework}\label{sec:vivas}

\begin{figure}[t]
  \begin{center}
    \includegraphics[width=0.6\linewidth]{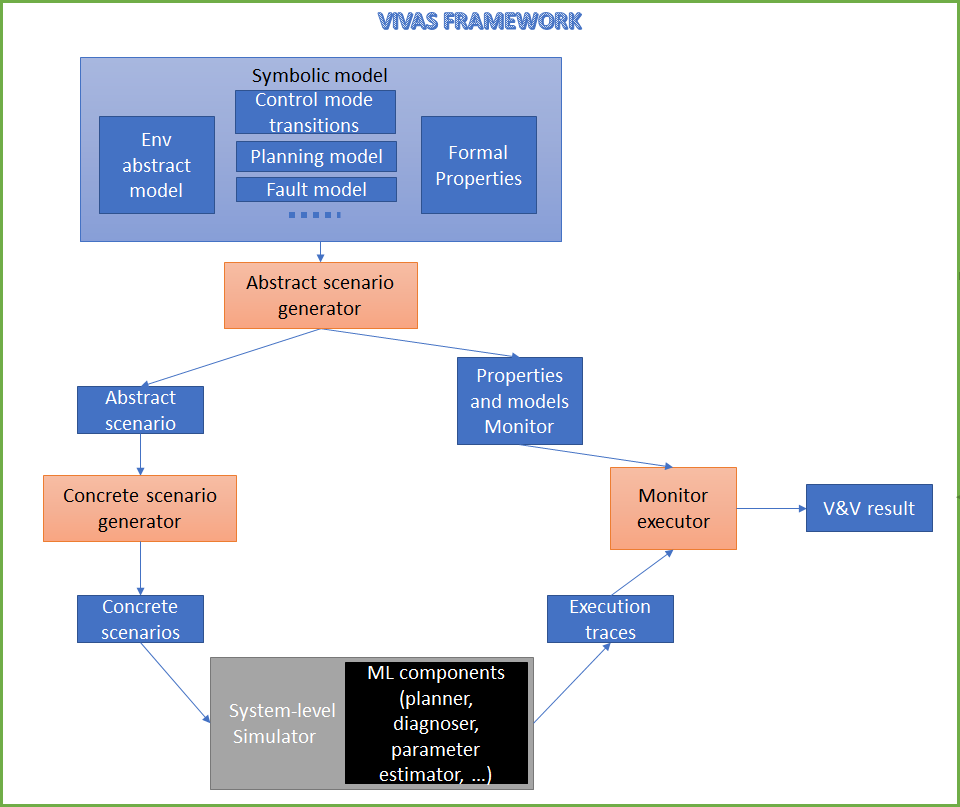}
    \caption{Top-level architecture of VIVAS framework. Blue boxes: artifacts, Orange boxes: code.
      \label{fig:vivasarch}}
  \end{center}
\end{figure}

VIVAS is a V\&V framework 
for generating test cases for autonomous systems (possibly using AI/ML components) 
via a combination of system-level simulation and symbolic model checking.
VIVAS makes use of formal,
symbolic models of the environment and system components to generate 
\emph{abstract test scenarios} for the autonomous system of interest using model checking techniques.
The abstract test scenarios are then instantiated by the
concretization of the abstract parameters to provide concrete scenarios to be
executed on a system-level simulator encompassing AI/ML models, to obtain
execution traces that are in turn analyzed by an automatically-generated
monitor. The output of the framework is a V\&V result consisting of coverage
statistics of the executed traces with respect to the symbolic models and
quantitative and qualitative information for each use case. The overview of the
architecture can be seen in Fig. \ref{fig:vivasarch}, which depicts the main
parts of the VIVAS framework. These are the 
abstract
scenario generator, the concrete scenario generator, the simulator, and the
executor monitor.

\paragraph{Abstract Scenario Generation.} Scenario generation is the first step of the approach. The starting point is a formal, symbolic model of the system, which provides an abstract view of both the environment and the components under test (including AI/ML parts). ML components are defined in a declarative manner, approximated in terms of input-output mapping. Abstract test scenarios are generated from the formal system model using symbolic model checking techniques by the abstract scenario generator. Abstract scenarios are defined as combinations of values of predicates describing interesting behaviors of the abstract system. From the technical point of view, each abstract scenario is encoded as a formal property that is expected to be violated by the system (i.e. a property specifying that ``the scenario cannot occur in the abstract system''). For each such property defined by the abstract scenario generator, a model checker will be executed on the system model, with the goal of finding a counterexample to the property. By construction, each such counterexample corresponds to an execution trace witnessing the realization of the abstract scenario of interest.

\paragraph{Concrete Scenario Generation.} Each of the traces produced by the
model checker is then refined into a (set of) concrete scenarios that can be
used to drive the system-level (concrete) simulator. Due to uncertainties and
abstractions in the abstract scenarios, a one-to-many mapping is defined where
a single abstract scenario can be instantiated to many, possibly infinite,
concrete scenarios. This is done by defining a mapping between the abstract
values and the set of concrete values that they represent and then
appropriately sampling from the sets. An example is that the abstract model
might discretize the time of day into dusk, dawn, midnight, and midday. The
concretizer then has to sample the actual time for the simulation. 

\paragraph{Simulation.} The task of the system-level simulator is to run a simulation of the target asset under the requested conditions, 
by configuring the system, its environment, and its inputs as specified in the concrete scenario produced by VIVAS.
Upon completion, the simulator provides the corresponding execution trace of the system, containing all the necessary details to evaluate the properties of interest.  

\paragraph{Execution Monitor.} Each concrete scenario produced is executed by the simulator, which generates a corresponding concrete execution trace. This trace is then used to determine whether: 
\begin{enumerate}
\item the concrete execution of the system satisfies the property of interest, and 

\item  the concrete execution of the system complies with the input abstract scenario (which defines the situation of interest for the current test). 
\end{enumerate}
This is done by formally evaluating the trace with a runtime monitor that is automatically generated from the formal specification of the property and the abstract system model. The trace evaluation can have four possible outcomes: 
\begin{enumerate}
\item The trace complies with the abstract scenario (defining the situation under test), and it also satisfies the property: the test execution is relevant and the test passes. 

\item  The trace complies with the abstract scenario, but it does not satisfy the property: this corresponds to a test failure on a relevant scenario, and it should be reported to the user. 

\item  The trace satisfies the property, but it does not comply with the abstract scenario: this corresponds to a (good) execution in an unexpected situation, in which some of the assumptions defining the scenario might be violated. This might be due to imprecisions/abstractions in the symbolic model and in the concretizer, which might prevent the realization of the abstract scenario under analysis. This situation might be reported to the user, as it might suggest that a revision/refinement of the symbolic model might be needed. 

\item  The trace violates the property and it does not comply with the abstract scenario: this corresponds to a test failure in an unexpected situation. Similarly to the above, it might be a warning that the symbolic model of the system is not precise enough to capture the situations of interest defined by the abstract scenario. 
\end{enumerate}

\paragraph{Abstract and concrete coverage.}

Ensuring an adequate level of coverage is one of the primary goals of a good set of tests. 
In VIVAS, coverage is defined with respect to a domain-specific notion of ``interesting situations'',
which are those that are (implicitly) defined by the possible combinations of values of predicates that are used by the abstract scenario generator to produce abstract traces.
By construction, therefore, VIVAS tries to enumerate abstract scenarios that ensure a 100\% degree of coverage of the \emph{abstract} situations of interest%
\footnote{Note that a 100\% degree of coverage might not be reached, either because some situations are not feasible already at the abstract level, or because the model checker cannot find a witness trace for the scenario specification within the given resource budget (time and/or memory).}.
Each abstract scenario is then refined into one or more concrete simulation inputs, leading to corresponding concrete simulation traces.
In order to determine the \emph{concrete} coverage (i.e., the degree of coverage of interesting situations at the concrete level), the VIVAS monitor analyzes the execution traces. It checks for compliance with the property of interest and the corresponding abstract scenario's specification (i.e., the ``interesting situation'') from which the concrete executions originate.

Only executions that satisfy the abstract scenario specification contribute to the coverage at the concrete level:
if an execution does not comply with its abstract specification,
it represents an unexpected situation from which no coverage information can be drawn%
\footnote{Note that in principle such a situation might still provide \emph{some} information (e.g. it might still cover a different but still interesting situation);
therefore, the test result is still reported to the user. 
However, determining this might not be obvious in general, and therefore we opted for the conservative choice of excluding the test from the computation of the degree of coverage in such cases.}.

\section{Autonomous Driving Application}
\label{sec:ads}
In order to apply the VIVAS methodology to ADS application, we instantiate various components of the VIVAS framework.
We choose \carla simulator as it is widely used in the automotive domain and it has a large community that provides various AI-based solutions for perception and control. 
We define an abstract model that focuses on highway scenarios where the ego is surrounded by other vehicles in various dynamic situations. 
In the following, we provide details about the different components.

\subsection{\carla Simulation Model and AI Components}
\carla \cite{carla} is a high-fidelity open-source simulator that provides a dynamic environment for the development, testing, and validation of AD systems.
It is written in C++ as a plugin for Unreal Engine. As a standalone package, it provides pre-defined maps with 3D meshes ranging from city roads with intersections to highways, to mimic real-world landscapes for the agents to drive in.
Various sensor models (cameras, Lidar, radar, GPS, IMU) are provided to gather the data from the environment.
The simulator includes many vehicle models, from small cars to large trucks, with different properties like mass, dynamics, and controls. 
A simulation is composed of (i) the CARLA Simulator that computes the physics
and renders the scene and all actor properties, (ii) client scripts written using a Python API, that allows control of the actors, sensors, and environmental conditions.

\paragraph{AI-based components.}
The \carla community through its leaderboard competition~\cite{carla-leaderboard} provides various state-of-the-art AI solutions for end-to-end autonomous driving. 
However, only a few of them provide the necessary code and well-trained models for their methodologies to be evaluated and built upon. We specifically tested Interfuser \cite{shao2022safetyenhanced}, TCP \cite{wu2022trajectoryguided}, and LAV \cite{chen2022learning}, all three currently in the top 5 of the leaderboard.
Within a few test runs of the AI agent provided with TCP, we noticed that the ego vehicle brakes to a standstill as soon as any other vehicle arrives next to it in its adjacent lane. We consider it too conservative of an autonomous behavior to test our verification methodology. 
The LAV agent on the other hand behaved well autonomously (in accordance with its overall score on the leaderboard) in terms of route completion and collision avoidance. However, it had an erratic behavior of changing lanes non-deterministically at scenario instantiation. It would require us to make ad-hoc changes to relative positions of the non-egos with respect to ego in every concrete scenario we generate.

We therefore chose the Interfuser agent as the AI system under test for our V\&V methodology. 
It is currently ranked 2 on the leaderboard (rank 1 among the open-source solutions). 
This solution primarily focuses on the safety of AD systems by generating interpretable semantic features of the environment through multi-model sensor fusion, for constraining the agent's low-level control actions in real-time within safe sets. 
The perception system processes the data gathered by 3 RGB cameras and one Lidar sensor.

\smallskip\noindent
All three AI agents mentioned above share the following main characteristics:
\begin{itemize}
\item The maximum driving speed is limited to $5\  m\!/\!s$, which is quite conservative for highway driving;
\item The ego always travels in its own lane: an external route for the ego to follow needs to be provided. It may change lanes only based on the waypoints of this route on the map. Hence, it never overtakes slow-moving cars in front of it in the same lane. Ego just follows them while maintaining a safe distance, or keeping a stand-still.
\item Standard rules of the road for overtaking only on the left (or the right) are not applied.
\end{itemize}

Note that our V\&V methodology is agnostic to the AI solution chosen for the simulator. Since the abstract test scenarios are generated from the symbolic model of the system, abstract coverage would be the same for different AI solutions, although the concrete coverage may vary. In future work, we will use our methodology to benchmark other AI solutions as well.

\subsection{Abstract Model and Coverage Criterion}
\label{sec:absmodel}

We specify our abstract model as a synchronous symbolic transition system written in the language of the nuXmv~\cite{nuxmv} model checker.
The model consists of 3 vehicles (one ``ego'' car, representing the autonomous system under test, and two other cars) moving on a highway with 3 lanes. 
The vehicles all drive in the same direction.
The ego is constrained to stay in the middle lane and tries to maintain a given cruise speed, braking when necessary to avoid collisions with other cars, and possibly accelerating to reach the target speed.
The other two ``non-ego'' cars can move freely on the highway, with arbitrary accelerations, braking, and lane change maneuvers (subject to physical constraints about min/max acceleration rates and speed limits, taken from publicly available online car databases), but are not allowed to crash into each other or the ego.
We use a discrete model of time, in which each transition of the system corresponds to a time-lapse of 1 second.
We use the theory of real arithmetic to encode the transition relation of the system, using mostly linear constraints to compute the updates to the speed and locations of the vehicles (thanks to the discretization of time). 
An excerpt of the symbolic model is shown in Fig.~\ref{fig:smv-model}. The module $\mathtt{Car}$ is shared by different non-ego vehicles. Different transition relations on speed, acceleration, and position need to hold when a non-ego changes lane (with $\mathtt{changing\_lane}$).
For the $\mathtt{Ego}$ module, we define the collision condition ($\mathtt{collision\_next}$) with non-ego vehicles ($\mathtt{car1}$ \& $\mathtt{car2}$, in this case). If True, the ego brakes with the $\mathtt{max\_braking}$ until it stops; else it continues with (or reach towards) its $\mathtt{target\_speed}$.
\begin{figure}[t]
  \centering
\begin{lstlisting}[language=smv]
MODULE Car(id)
IVAR acceleration : real;
VAR pos : real;
    lane : 0 .. MAX_LANE;
    speed : real;
DEFINE changing_lane := next(lane) != lane;
TRANS
  changing_lane -> (speed <= max_lane_change_speed & 
                    next(speed) <= max_lane_change_speed);
TRANS
  changing_lane -> (acceleration <= max_lane_change_acceleration & 
                    acceleration >= (- max_lane_change_braking));
TRANS
  next(speed) = max(speed + acceleration * TIME_STEP, 0);
TRANS
  changing_lane ?
     (next(pos) = pos + (speed + next(speed)) / 2 * TIME_STEP * 0.95) :
     (next(pos) = pos + (speed + next(speed)) / 2 * TIME_STEP);
TRANS
  (next(lane) = lane) | (next(lane) = lane + 1) | (next(lane) = lane - 1);


MODULE Ego(car1, car2)
-- VAR declarations...
DEFINE
  time_to_stop := speed / (-MAX_BRAKING);
  collision_next := (car1.lane = lane & car1.pos >= pos & speed > 0 & 
                       (car1.pos - pos) / speed <= time_to_stop) | 
                    (car2.lane = lane & car2.pos >= pos & speed > 0 & 
                       (car2.pos - pos) / speed <= time_to_stop);
TRANS
  collision_next ?
  (acceleration = MAX_BRAKING & target_speed = 0) :
  (target_speed = EGO_CRUISE_SPEED &
  ((speed < target_speed) -> 
                (next(speed) = min(target_speed,
                                   speed + MAX_ACCELERATION * TIME_STEP))));

INVAR -- the cars do not crash into each other on purpose
  ((abs(pos - car1.pos) > SAFE_DISTANCE) | (lane != car1.lane)) &
  ((abs(pos - car2.pos) > SAFE_DISTANCE) | (lane != car2.lane)) &
  ((abs(car1.pos - car2.pos) > SAFE_DISTANCE) | (car1.lane != car2.lane));
\end{lstlisting}  
  \caption{Excerpt of the nuXmv code for the abstract model.
    \label{fig:smv-model}}
\end{figure}

\begin{figure}[t]
  \centering
  \includegraphics[width=0.25\linewidth]{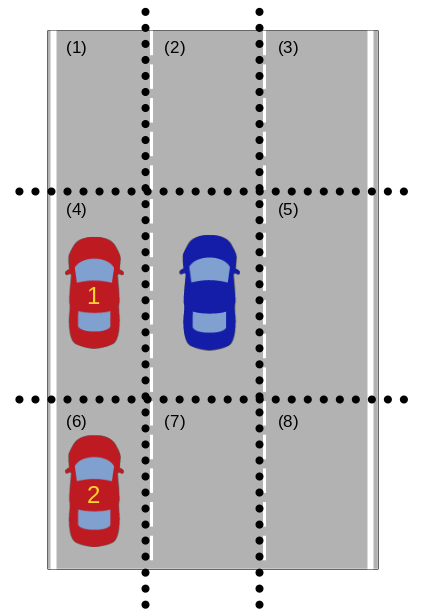}
  \hspace{5em}
  \includegraphics[width=0.25\linewidth]{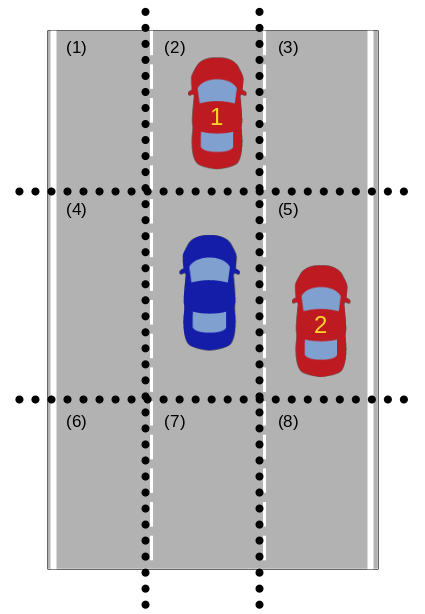}
  \caption{Example traffic situations for constructing abstract scenarios,
    specifying positions of non-ego cars (in red) in terms of the occupation of cells of an abstract 3x3 grid centered on the ego car (in blue).
    \label{fig:scenarios}}
\end{figure}

In order to enumerate abstract scenarios encoding potentially-interesting traffic situations, 
we define for each non-ego car a set of predicates specifying its position relative to the ego, in terms of occupation of cells of an abstract 3x3 ``grid'' centered on the ego. 
Examples of the possible configurations that can be expressed in this way are shown in Fig.~\ref{fig:scenarios}.
We then define an \emph{abstract scenario} as a combination of constraints about the different positions of the non-ego cars on the grid at different points in time.
More specifically, each abstract scenario is specified as an LTL property of the following form:
\begin{equation}\label{scenario-spec}
  \begin{split}
    \neg \mathbf{F}(& \text{car1\_grid\_pos = CELL\_A1} \land \text{car2\_grid\_pos = CELL\_A2} \land \\ & \mathbf{X}(\mathbf{F}(\text{car1\_grid\_pos = CELL\_B1} \land \text{car2\_grid\_pos = CELL\_B2}))),
  \end{split}
\end{equation}
\noindent (where $\mathrm{car\mathit{i}\_grid\_pos}$ encodes the position of the $i$-th non-ego car in the grid and $\mathrm{CELL\_*}$ represent possible target positions for the cars.)
By asking the model checker to find a counterexample to Eq. \ref{scenario-spec}, 
we generate traces in which the non-ego cars first reach the configuration with $\mathtt{car1}$ in position $\mathrm{A1}$ and $\mathtt{car2}$ in position $\mathrm{A2}$, and then subsequently move to the configuration with $\mathtt{car1}$ in position $\mathrm{B1}$ and $\mathtt{car2}$ in position $\mathrm{B2}$, 
performing the necessary maneuvers while avoiding collisions with each other or with the ego.

The space of scenarios that is being explored therefore consists of all the possible combinations of transitions from configurations of the non-ego cars in terms of their position in the grid defined above. Enumerating all of them would give 4096 scenarios.
We define our coverage criterion by selecting a subset of \emph{abstract scenarios of interest}, 
consisting of various combinations of the traffic situations that can be modeled by positioning the non-ego cars in the grid around the ego.
In total, for the experiments, we defined 144 such interesting scenarios.

\subsection{VIVAS Interface with \carla}
\label{sec:vivasinterfacecarla}
In
order to generate a scenario for the \carla simulator, the abstract
counterexample trace generated by the model checker is parsed for
relevant information to be fed as input to the simulator. As an
interface to the simulator, we used the \carla module
\scenariorunner. 
This provides a Python interface
to specify the routes for the ego as well as complex
traffic scenarios by defining the behavior of the non-ego
agent(s). \scenariorunner also allows for running \carla on a
specified map at a particular location, while the user is allowed to
implement their own AI-based ego agent. Every state of the abstract
scenario trace is concretized into the corresponding behavior of
every non-ego agent. Each behavior is then specified in Python to
generate a behavior tree for each corresponding non-ego vehicle. The
behavior trees of all the non-egos present in the environment are then
run in parallel during the simulation.

We first parse the initial coordinates and lanes of all the non-egos relative to the ego to instantiate them on the map. In each behavior of a behavior tree, the corresponding non-ego has to drive at a certain speed for a certain distance, following the waypoints given by the map on the same lane it is instantiated on.
Although the duration of each state transition in the abstract trace is 1 second, the non-egos may take longer to drive that particular distance in the \carla simulator, due to potential mismatches between the symbolic model and the simulator models.
In case the vehicle stands still for $n$ states in the abstract trace, it stands still for $n$ seconds in the concrete scenario once it comes to a halt.

In the symbolic model, lane changes occur in one time step, with zero lateral distance traveled (since lanes have no width in the symbolic world). However, we constrain the successive lane changes of the same car to be $N$ steps apart%
\footnote{We used $N=6$ in our experiments.}
to model the fact that a lane change is not instantaneous overall. 
To concretize this particular state transition, the non-ego transverses $9m$  while changing  lanes, with this behavior terminating after traveling a total distance of $12m$ for the next behavior in the tree to be instantiated. Below these values, lane changes were not possible in \carla at the speed ranges the vehicles drive in our scenario. Note that similar to the symbolic model, a non-ego can change only one lane at a time, with inputs $\mathrm{\{left, right\}}$ meaning change lane to the left or to the right.
\begin{figure}[t]
	\centering
	\includegraphics[width=0.9\linewidth]{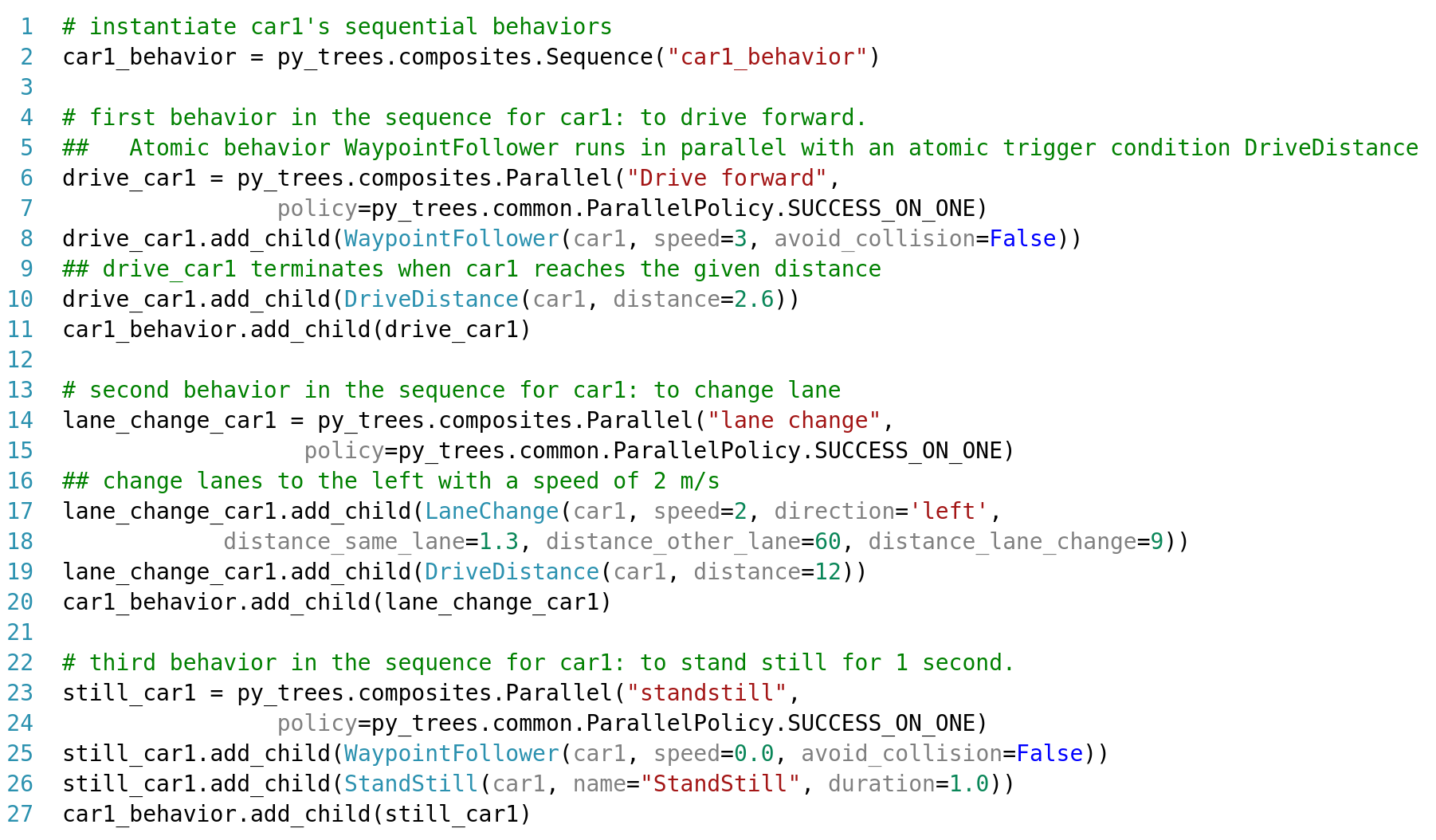}
	\hspace{2em}
	\caption{Excerpt of automatically generated Scenario runner code (in Python3) for a non-ego (car1) behavior. Comments (in green) explain the behavior tree.
		\label{fig:SRcode}}
\end{figure}

We leverage the behavior library of \scenariorunner to write these atomic behaviors and trigger conditions. An example behavior tree for one non-ego (car1) with all the above explained three behaviors is shown in Fig.~\ref{fig:SRcode}. Here, lines 6-11: drive straight forward for 2.6 $m$, with a speed of 3 $m/s$; lines 14-20: perform a lane change to the left, with a speed of 2 $m/s$, driving a total of 12 $m$ within which 9 $m$ is the distance traveled while changing lanes; lines 23-37: stand still for 1 second.
We do not need to extract ego's behavior from the abstract trace, since it is expected to make decisions autonomously in the simulator. Only initial spawn position and destination need to be extracted for the AI-based agent to follow the route.

The concrete simulation traces are then mapped back to the abstract trace to measure coverage, 
to check if the same sequence of scenes was encountered in the concrete scenarios or not. In particular, a predicate map is defined to map the absolute positions of the non-egos in the concrete simulation trace to the abstract 3x3 grid shown in Fig. \ref{fig:scenarios}. Here, we show examples of mapping the positions of non-egos to the cell locations 1,4 and 8 of the abstract grid:
\begin{equation}
  \label{eq:gridcells}
  \begin{aligned}
    & \text{\bf CELL\_1}: (\text{car\_i.lane} < \text{ego.lane}) \land (4 \le |\text{car\_i.pos - ego.pos}| \le 24) \land (\text{car\_i.pos} > \text{ego.pos}) \\
    & \text{\bf CELL\_4}: (\text{car\_i.lane} < \text{ego.lane}) \land (|\text{car\_i.pos} - \text{ego.pos}| \le 10) \\
    & \text{\bf CELL\_8}: (\text{car\_i.lane} > \text{ego.lane})  \land (4 \le |\text{car\_i.pos} - \text{ego.pos}| \le 24)  \land (\text{car\_i.pos} < \text{ego.pos}),
  \end{aligned}
\end{equation}
\noindent 
where car\_i.pos is the longitudinal position (in meters) of the $i$-th car in the simulation trace (and similarly for ego.pos). In this way, we define the boundaries of the cells on the abstract grid.

\subsection{Monitoring of Properties}

As described above, the monitor component of VIVAS is used to determine whether the concrete system (simulator) satisfies the system-level formal specification.
For the automotive application, the simulation output traces include sequences of all states and actions executed by the ego vehicle, along with the time evolution of other observable parameters, which must be checked for property satisfaction/violation.
In this study, we primarily need to check whether the ego vehicle crashes with another vehicle in the environment.
Since the ego always travels in its own lane, we limit the check for the case when the ego crashes with any non-ego in front of it in its own lane.
We do this by leveraging the continuous data stream from the collision sensor mounted on the ego. 
The monitor is currently hard-coded for monitoring specifically the output of this sensor, i.e., it checks whether the ego crashes or not at any time step in the simulation trace.
Along with the satisfaction/violation of this property, 
the positions of non-egos in the simulation trace are mapped back to the abstract grid,
to measure the degree of concrete coverage as described in \S\ref{sec:vivas}.

In the future, we plan to use a runtime monitor based on NuRV~\cite{nurv,CimattiTT19}, to check standard LTL properties on ego behavior, e.g., if ego brakes within $n$ time-steps as soon as any non-ego comes in front within its safe driving distance, or if the lane change of another vehicle is detected by the perception component of the ego within $m$ time steps.
\section{Results}
\label{sec:results}

In this section, we report on our experimental evaluation of our instantiation of VIVAS for the ADS application using the \carla simulator.
We first describe the experimental setup in \S\ref{sec:experiments-setup}, including the choice of parameters for the vehicles and environment in the symbolic model and in the \carla simulator, necessary to generate meaningful scenarios.
We then present the results of the evaluation in \S\ref{sec:experiments-results} and discuss them in \S\ref{sec:experiments-discussion}.

We ran the experiments on an Intel i7 with NVIDIA GeForce RTX 2080 8GB GPU. These are the minimum hardware requirements to run the simulations on the \carla simulator with AI models.
All the experiments take roughly 22 hours to complete. 
We used a timeout of 200 seconds for each abstract scenario generation. This timeout was never reached by the model checker during our experiments: on average, model checking took less than 10 seconds for each instance.
Rather, the performance bottleneck turned out to be the time to instantiate a scenario in \carla and perform the simulations, which took 3 minutes on average.

The code and data necessary for reproducing our experiments are available at 
\url{https://es-static.fbk.eu/people/sgoyal/fmas23}.

\subsection{Experimental setup}
\label{sec:experiments-setup}

At the level of the symbolic model, we define here some fixed parameters for simple, but meaningful scenario generation:
\begin{equation}
	\begin{aligned}
		&\text{ego\_cruise\_speed} &=& \ 5 \ &(m/s) \\
		&\text{non\_ego\_speed} &=& \ [0, \ 12] \ &(m/s) \\
		&\text{max\_acceleration} &=& \ 5.6 \ &(m/s^2) \\
		&\text{max\_braking} &=& \ -4.6 \ &(m/s^2) \\
		&\text{safe\_distance} &=& \ 7 \ &(m) \\
		&\text{lanes} &=& \ \{\text{left, \ center, \ right}\} \\
	\end{aligned}
\end{equation}

All the scenarios that we generate consist of 2 non-ego agents and one ego agent, all of which start from the same longitudinal position $x=0$, with ego in the center lane and 2 non-egos on each side of it.
Note that fixing the initial positions would not make a difference to the abstract scenario generation since the acceleration, braking, and speed for the non-egos are picked non-deterministically by the model checker for every time step, while respecting the above bounds.
Since the AI-based agent we use in \carla can only drive at around $5 \ m/s$, we limit the ego agent cruising speed to the same.
All the agents start from $0\ m/s$, with ego reaching its cruising speed with $\mathrm{max\_acceleration}$. To avoid collisions, it brakes with $\mathrm{max\_braking}$ to maintain at minimum the $\mathrm{safe\_distance}$ with all the non-egos. 

To improve the robustness of abstract scenario generation, we reduce the size of each cell in the abstract 3x3 grid in the symbolic model (see Fig. \ref{fig:scenarios}) by 3m in each direction, compared to the grid we use for evaluation of the coverage on the simulator. 
The lower values of the relative distances here are chosen to specifically create situations where non-egos stay close to ego and challenge its perception and control system with their braking and lane-changing maneuvers.

To mimic the symbolic environment model, we instantiate the \carla simulator on a section of a highway of Town06, with 5 straight lanes, with ego positioned on the center lane and two non-egos on each lane next to it, corresponding to the symbolic model. Left-most and right-most lanes are not used.
To compensate for the mismatch between the vehicle dynamics in the symbolic model and \carla simulator, we concretize the initial positions of the non-egos at:
\begin{equation}
	\label{eq:xpos}
	x = \{-3.5, 0, 3.5\} \ m
\end{equation}

i.e., the non-egos start at 3.5 $m$ behind, same level, and 3.5 $m$ ahead of the ego in their respective lanes in different simulation runs. All the vehicles start from 0 $m/s$, as parsed from the abstract trace. In future work, we could also concretize further for the simulations with one non-ego ahead and the other one behind the ego, to check if it extends the coverage results. 

\carla provides the possibility to change weather conditions (e.g., rainy, cloudy, night, etc.) at the beginning of simulation runs. We perform all the simulations in "clear noon" setting, for the ego's perception components to operate in the least challenging conditions. In future work, we will evaluate the AI solution in different weather conditions.

\subsection{Evaluation}
\label{sec:experiments-results}
The model checker produces a total of 144 abstract scenarios based on the coverage criteria given in \S\ref{sec:absmodel}. Each abstract scenario is concretized into three concrete scenarios, by varying the initial positions of the non-egos according to Eq. \ref{eq:xpos}, which gives us $144*3=432$ concrete scenario outputs from the simulator. The evaluation results are shown in Table \ref{tab:Results}. The columns have the following meanings:

\begin{description}
	\item[Coverage OK:] Each point of the grid of coverage criteria represents a scenario with a fixed order of scenes. The concrete simulation run passes ("OK") if the abstract scenario generated by the model checker could indeed be generated on the simulator as well. 
	
	\item[Property FAIL:] The system-level property fails if the autonomous ego agent collides with at least one non-ego in front. We do not take into account the situations where non-egos crash into each other or hit the ego from behind.
	
	\item[Coverage OK $\cap$ Property FAIL:] the intersection of the above two conditions. These are the set of "interesting" cases (along with the other cases where property failed), where the coverage criteria passed, but the ego crashed with a non-ego in front.
	
	\item[Set Union:] Combines the results for all concrete scenarios with respect to the abstract scenarios.
\end{description}

\begin{table}[t]
	\begin{center}
          \begin{small}
            \caption{Coverage Results.
              \label{tab:Results}}
            \begin{tabular}{|cc|c|c|c|}
              \hline
              \multicolumn{1}{|c|}{\multirow{2}{*}{\bf Non-ego position}} &
                                                                            \multirow{2}{*}{\bf Total Scenarios} &
                                                                                                                   \multirow{2}{*}{\bf Coverage OK} &
                                                                                                                                                      \multirow{2}{*}{\bf Property FAIL} &
                                                                                                                                                                                           \multirow{2}{*}{\bf Coverage OK $\cap$ Property FAIL} \\
              \multicolumn{1}{|c|}{}           &       &             &             &             \\ \hline
              \multicolumn{1}{|c|}{3.5 m}      & 144   & 68          & 3           & 2           \\ \hline
              \multicolumn{1}{|c|}{0}          & 144   & 73          & 17          & 4           \\ \hline
              \multicolumn{1}{|c|}{- 3.5 m}    & 144   & 45          & 54          & 19          \\ \hline
              \multicolumn{2}{|c|}{\textbf{Set Union}} & \textbf{81} & \textbf{61} & \textbf{25} \\ \hline
            \end{tabular}
          \end{small}
	\end{center}
\end{table}


\subsection{Analysis}
\label{sec:experiments-discussion}

All intended 144 abstract counterexamples could be generated by the model checker, meaning that the configurations we defined for the coverage did not violate any constraints in the symbolic model. As we see from the obtained results, not all the abstract scenarios generated by the model checker could be covered in the simulator. We could cover a total of only 81 out of 144 scenarios.
This is primarily due to the mismatch in (a) behavior models and (b) vehicle dynamics, between the symbolic model and the simulator.

\paragraph{Behavior model mismatch.} Since the ego is based on AI models, its non-deterministic behavior is not fully represented in the symbolic model. Ego speed is always varying within +-1 $m/s$ compared to the constant cruising speed in the symbolic model. 
The bounds of cells in the predicate map are defined for the relative position of the non-egos with respect to ego, as specified in Eq. \ref{eq:gridcells}. Hence, in some cases, the non-egos can not reach the required region within these bounds during the scenario, since the ego is traveling too fast or slow. 
In principle, we could overcome this by conditioning the non-egos' behavior to the ego's in terms of the distance traveled relative to the ego instead of the absolute distance on the lane, while translating the abstract scenario to the concrete one. However, no such atomic behaviors or conditions yet exist in \scenariorunner. This could be included in one of the future works.

\paragraph{Vehicle dynamics mismatch.} The vehicle dynamics models
          are based on OEM data, hard-coded in the simulator. The
          physical constraints for maximum acceleration and braking in
          the symbolic model are instead generic. 
 However, we found a big mismatch during the simulations, with maximum acceleration values of vehicle models in \carla reaching as high as 12 $m/s^2$, and maximum braking below -15 $m/s^2$.
Since these parameter values also vary from vehicle to vehicle in \carla, we even came across situations where non-egos crashed into each other while changing lanes.

\begin{figure}[t]
	\centering
	\includegraphics[width=0.2\linewidth]{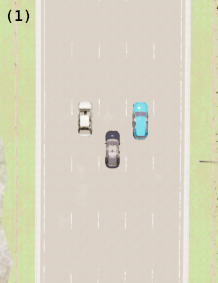}
	\hspace{2em}
	\includegraphics[width=0.2\linewidth]{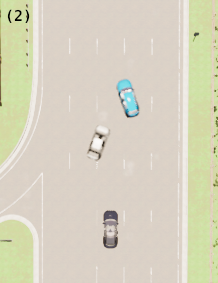}	\hspace{2em}
	\includegraphics[width=0.2\linewidth]{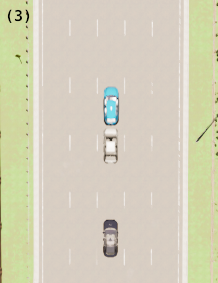}
	\\[2em]
	\includegraphics[width=0.2\linewidth]{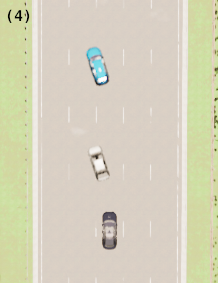}
	\hspace{2em}
	\includegraphics[width=0.2\linewidth]{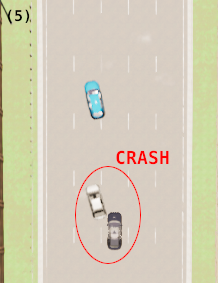}
	\hspace{2em}
	\includegraphics[width=0.2\linewidth]{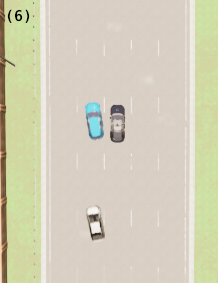}
	\caption{Scenario defined in Eq. \ref{eq:int-scenario-spec}, with Coverage OK $\cap$ Property FAIL. Ego crashing with a non-ego in front (scene 5).
		\label{fig:crash}}
\end{figure}

\paragraph{Interesting scenarios.}

Even though we could not cover all the abstract scenarios in the
concrete simulator, there were 61 scenarios where AI-based ego
collided with a non-ego agent in front. In particular, we obtained 25 interesting
scenarios that met their abstract specification, but with ego crashing
into a non-ego in front. Fig. \ref{fig:crash} shows 6 scenes (in temporal order of 1-6) extracted from one such scenario. This corresponds to the LTL
property specified below in Eq. \ref{eq:int-scenario-spec} (with reference to Eq. \ref{scenario-spec}):
\begin{equation}\label{eq:int-scenario-spec}
	\begin{split}
		\neg \mathbf{F}(& \text{car1\_grid\_pos = 2} \land \text{car2\_grid\_pos = 2} \land \mathbf{X}(\mathbf{F}(\text{car1\_grid\_pos = 6} \land \text{car2\_grid\_pos = 4})))
	\end{split}
\end{equation}

\begin{figure}[t]
	\centering
	\includegraphics[width=0.53\linewidth]{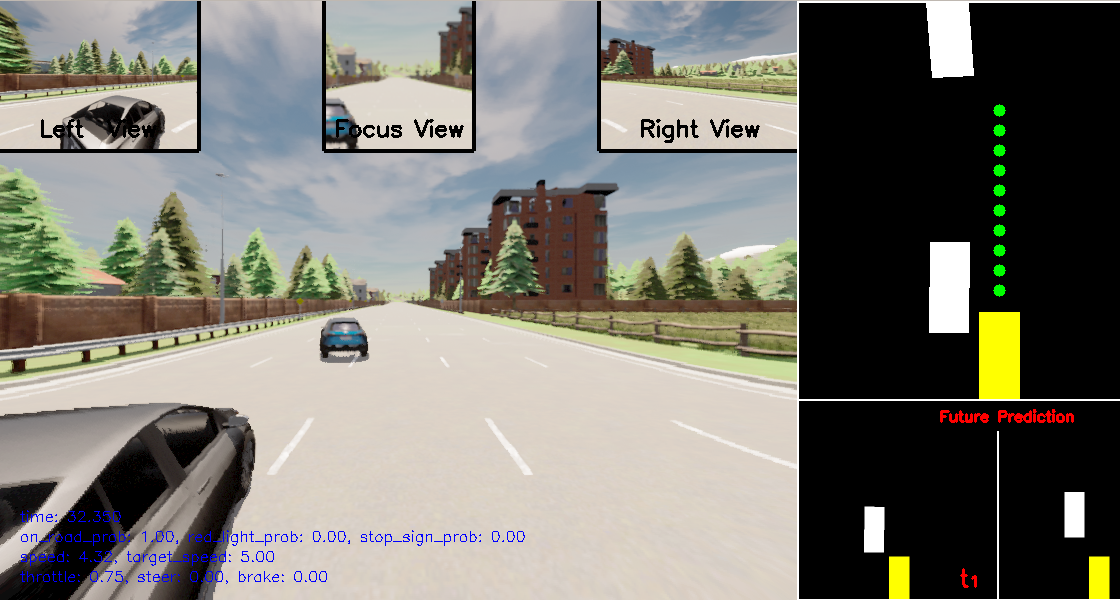}
	\hspace{1em}
	\includegraphics[width=0.42\linewidth]{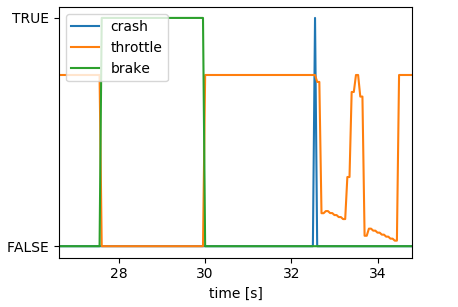}
	\caption{Ego collision; (left): Different camera views of AI-based ego, with the perception component's output; (right): Plot showing ego's throttle and braking values during the collision (in blue).
		\label{fig:crashfrontview}}
\end{figure}

\noindent
Here, the grid positions correspond to the cell numbers mentioned in Fig. \ref{fig:scenarios} for the abstract grid space. We now describe the scenario in Fig. \ref{fig:crash}.
\begin{itemize}
	\item All the agents are initialized at 0 $m/s$, with car1 on the left of ego, and car2 to its right, with both cars starting 3.5 $m$ ahead of ego (scene 1). Here, ``$\text{car1\_grid\_pos = 4} \land \text{car2\_grid\_pos = 5}$'' on the abstract grid.
	\item The non-egos travel faster than the ego to change lanes in front of it (scene 2), and end up in the configuration (scene 3) where the first predicate, ``$\text{car1\_grid\_pos = 2} \land \text{car2\_grid\_pos = 2}$'' is satisfied.
	\item The non-egos then change lanes to the left slowly (scene 4), when the ego crashes with car1 (scene 5), even when the car1 has still not completed the lane change.
	
	Fig.~\ref{fig:crashfrontview} (left) shows the front view of the AI agent at the instant of crashing with non-ego. The real-time telemetry shows that ego's brake = 0 and throttle = 0.75. The perception component here seems to mis-detect the actual position of car1. The corresponding output from the simulation trace is shown in the right plot, with ego's throttle = True (and brake = False) for more than 2 seconds leading to the collision.
	\item The ego keeps driving forward when the second predicate, ``$\text{car1\_grid\_pos = 6} \land \text{car2\_grid\_pos = 4}$'' is satisfied (scene 6).
\end{itemize}

\section{Conclusions}
\label{sec:conclusions}

This paper showed the application of system-level simulation-based
verification of ADS using formal methods to generate abstract
scenarios. The verification toolchain includes nuXmv for model
checking and generating abstract scenarios, \carla for simulating the
ego behaviors in concrete scenarios, and mappings from abstract to
concrete scenarios and back. We presented an abstract model of the
system and a coverage criterion that allows the automated generation
of abstract scenarios with model checking. The generated abstract
scenarios cover different sequences of traffic scenes that are
relevant to test the reaction of the ego's behavior to see if it
avoids crashing into other cars. The simulation with \carla of the
corresponding concrete scenarios showed various crashes caused by the
ego, although not all simulations reproduce the expected abstract
scenario. Inspecting some of the simulations reporting a crash in a
covered abstract scenario confirms that the ego behavior is indeed
buggy and this is probably due to the AI-based perception component.    

During this study, we gained many insights that may lead to some
interesting future research directions. These include more efficient
techniques to generate abstract scenarios for minimizing the number of
model checking runs needed to achieve a certain coverage level; the
integration of effective sampling techniques that synthesize various
simulation parameters for the same abstract scenario; extending the
abstract model by incorporating uncertainty in the ego behavior or
a more precise representation of the continuous-time behavior with
timed or hybrid version of SMV \cite{CimattiGMRT19,CimattiGMT15}; finally,
enhancing the concrete scenario specification with conditional
behaviors of non-ego vehicles that react to the choices of the ego.

\newpage

\nocite{*}
\bibliographystyle{eptcs}
\bibliography{refs}


\end{document}